\documentclass[aps,preprint]{revtex4-1}
\usepackage{bm}
\usepackage{amsfonts}
\usepackage{amsmath}
\usepackage{amssymb}
\usepackage{graphicx}

\begin{document}

\title{Neutrino emissivity of anisotropic neutron superfluids}
\author{L. B. Leinson}

\affiliation{Institute of Terrestrial Magnetism, Ionosphere and Radio Wave Propagation
RAS (IZMIRAN), 142190 Troitsk, Moscow, Russia}

\keywords{Neutron star, Neutrino radiation, Superfluidity}
\pacs{26.60.-c, 74.20.Fg, 26.30.Jk}

\begin{abstract}
We examine the influence of the anisotropy of the superfluid energy gap and
residual Fermi-liquid interactions in the triplet-correlated neutron liquid
onto neutrino energy losses through neutral weak currents. The neutrino-pair
emission caused by the pair breaking and formation processes and by the
spin-wave decays is considered for the case of the $^{3}P_{2}$ pairing in
the state with $m_{j}=0$. The simple analytical formulae are obtained. A
comparison with the previous results of the average-angle approach shows
that the gap anisotropy leads to quenching of the neutrino emissivity caused
by the pair recombination processes on about 15\% and to substantial
suppression of the spin-wave decays. Residual particle-hole interactions
increase the energy losses in both the channels on about 5\%.
\end{abstract}

\maketitle

\startpage{1}

\section{Introduction}

Neutrino emission from a superfluid neutron liquid is currently thought to
be the dominant cooling mechanism of the baryon matter, for some ranges of
the temperature and/or matter density. One of the mechanisms leading the
neutron star cooling consists on the recombination of thermally excited
baryon BCS pairs into the condensate with emission of neutrino pairs via
neutral weak currents \cite{FRS76}. It is generally accepted that, for
temperatures near the associated superfluid critical temperatures $T_{c}$,
emission from pair breaking and formation (PBF) processes dominates the
neutrino emissivities in many cases. This idea has become the basis of the
minimal cooling paradigm for superfluid neutron stars \cite{Page04,Page09}.

According to modern investigations \cite{LP06,KV08,L08,L09,SR}, the neutrino
radiation caused by the pairing of baryons into singlet state is strongly
suppressed, and the dominant neutrino emission, at the long-cooling epoch,
occurs from the triplet-correlated condensate of superfluid neutrons which,
as expected, exists in the superdense core of the star. Neutrino energy
losses owing to the triplet PBF processes have been initially derived in
Ref. \cite{YKL}, ignoring the anomalous weak interactions. The
self-consistent approach to this problem was developed in Refs. \cite%
{L10a,L10c,L11c}, where the anomalous effective vertices are found as the
solution of the Dyson's equations in the ladder approximation. Poles of the
found analytical solutions \cite{L10b,L12} indicate the existence of the
undamped collective oscillations of the order parameter in the superfluid $%
^{3}$P$_{2}$ condensate of neutrons with $m_{j}=0$. Some of the collective
excitations with the excitation energy smaller than the superfluid energy
gap are able to decay into neutrino pairs. This is, for example, the
"so-called" normal-flapping mode which represents the nonunitary excitation
with a nonzero average spin expectation value and can be termed by spin
wave. The neutrino decay of such spin waves could be important for
thermally-emitting neutron stars, presumably cooling through the combination
of neutrino emission from the interior and photon cooling from the surface,
the latter is responsible for their observed thermal emissions \cite{PZ02}.

Exact solutions of the vertex equations are complicated by the anisotropy of
the triplet order parameter. Therefore the preliminary analysis of the
collective oscillations and the neutrino energy losses from the $^{3}$P$_{2}$
superfluid neutron liquid has been performed in the average-angle
approximation replacing the anisotropic energy gap in the quasiparticle
energy by its value averaged over angles \cite{L10b,L11a,L11b}. The
spin-wave energy, as found in this approach, is proportional to the
temperature-dependent energy gap amplitude $\Delta \left( T\right) $.

However the more rigorous analysis with taking into account of the gap
anisotropy has shown that the average-angle approximation is not justified
for examination of the collective oscillations of the superfluid condensate.
It was found \cite{L12} that the gap anisotropy leads to a strong decreasing
of the level energy along with a lowering of the temperature. One can expect
that this is to suppress substantially the neutrino energy losses caused by
the spin wave decays because the rate of neutrino losses is strongly
dependent on the wave energy.

In this paper we examine the effect of the anisotropy of the superfluid
energy gap and the residual particle-hole interactions onto the neutrino
energy losses from the $^{3}$P$_{2}$ superfluid neutron liquid with $m_{j}=0$%
. We consider the neutrino emission caused by the PBF processes and by the
spin-wave decays (SWD).

The paper is organized as follows. The next section contains some
preliminary notes. We discuss the effective strong interactions in the
superdense neutron matter, the order parameter for the triplet
pair-correlated system with strong interactions and introduce the notation
used below. In Sec. III, we present the general expression for neutrino
emissivity of the medium through neutral weak currents in terms of \ the
imaginary part of the current-current correlator. In Sec. IV, we examine the
anomalous vertices with taking into account of both the Fermi-liquid effects
and the anisotropy of the superfluid energy gap for the case of $^{3}$P$_{2} 
$ pairing with $m_{j}=0$, as is adopted in the minimal cooling paradigm \cite%
{Page09}. We also discuss the self-consistent response of the superfluid
neutron liquid onto external neutrino field. Finally, in Sec. V, we evaluate
the self-consistent neutrino energy losses from the PBF and SWD processes
and compare them with the neutrino losses caused by modified urca processes
and nn-bremsstrahlung. Section VI contains a short summary of our findings
and the conclusion. A short Appendix contains some transformation of the
previous results.

In this work we use the standard model of weak interactions, the system of
units $\hbar =c=1,$ and the Boltzmann constant $k_{B}=1$.

\section{Preliminary notes and notation}

The order parameter, $\hat{D}\equiv D_{\alpha \beta }$, arising owing to
triplet pairing of quasiparticles in a degenerate Fermi system, represents a 
$2\times 2$ symmetric matrix in spin space, $\left( \alpha ,\beta =\uparrow
,\downarrow \right) $. Near the Fermi surface this matrix can be written as
(see e.g. \cite{Ketterson}) 
\begin{equation}
\hat{D}\left( \mathbf{n}\right) =\Delta \mathbf{\bar{b}}\left( \mathbf{n}%
\right) \cdot \bm{\hat{\sigma}}\hat{g}~,  \label{Dn}
\end{equation}%
where $\bm{\hat{\sigma}}=\left( \hat{\sigma}_{1},\hat{\sigma}_{2},\hat{\sigma%
}_{3}\right) $ are Pauli spin matrices, and $\hat{g}=i\hat{\sigma}_{2}$.

The spin-angle structure of the triplet condensate is defined by a vector $%
\mathbf{\bar{b}}\left( \mathbf{n}\right) $ in spin space which depends on
the direction of the quasiparticle momentum $\mathbf{p}$. The angular
dependence of the order parameter is represented by the unit vector $\mathbf{%
n=p}/p$ which defines the polar angles $\left( \theta ,\varphi \right) $ on
the Fermi surface, 
\begin{equation}
n_{1}=\sin \theta \ \cos \varphi ,\ \ \ n_{2}=\sin \theta \ \sin \varphi ,\
\ \ n_{3}=\cos \theta .  \label{n}
\end{equation}%
We assume that the (temperature-dependent) gap amplitude $\Delta $ is a
complex constant (on the Fermi surface), and $\mathbf{\bar{b}}\left( \mathbf{%
n}\right) $ is a real vector which we normalize by the condition 
\begin{equation}
\left\langle \bar{b}^{2}\left( \mathbf{n}\right) \right\rangle =1~.
\label{Norm}
\end{equation}%
Hereafter we use the angle brackets to denote angle averages, 
\begin{equation}
\left\langle ...\right\rangle \equiv \frac{1}{4\pi }\int d\mathbf{n}\cdot
\cdot \cdot =\frac{1}{2}\int_{-1}^{1}dn_{3}\int_{0}^{2\pi }\frac{d\varphi }{%
2\pi }\cdot \cdot \cdot .  \label{av}
\end{equation}

The triplet pairing leads to the energy gap, $\Delta \bar{b}\left( \mathbf{n}%
\right) $, in the quasiparticle spectrum which is in general anisotropic. We
are mostly interested in the values of quasiparticle momenta $\mathbf{p}$
near the Fermi surface $p\simeq p_{F}$,\ where the quasiparticle energy is
given by%
\begin{equation}
E_{\mathbf{p}}=\sqrt{\varepsilon _{p}^{2}+\Delta ^{2}\bar{b}^{2}\left( 
\mathbf{n}\right) }~,  \label{Ep}
\end{equation}%
with%
\begin{equation}
\varepsilon _{p}\simeq V_{F}(p-p_{f}),  \label{ksi}
\end{equation}%
and $V_{F}\ll 1$ is the Fermi velocity of the nonrelativistic neutrons. $%
\allowbreak \allowbreak $Here the fact is used that, in the absence of
external fields, the gap amplitude $\Delta $ is real.

The spin-orbit interaction among quasiparticles is known to dominate in the
nucleon matter of a high density \cite{Tamagaki,Takatsuka} with the most
attractive channel of interactions in the $^{3}$P$_{2}$ state with $%
s=1,j=2,l=1$. In this case the order parameter in the superfluid system can
be constructed with the aid of the set of mutually orthogonal complex
vectors $\mathbf{b}_{m_{j}}\left( \mathbf{n}\right) $ which generate
standard spin-angle functions of the total angular momentum $j=2$ and $%
m_{j}=0,\pm 1,\pm 2$, so that 
\begin{equation}
\mathbf{b}_{m_{j}}(\mathbf{n})\hat{\bm{\sigma}}\hat{g}\equiv
\sum_{m_{s}+m_{l}=m_{j}}\left( \frac{1}{2}\frac{1}{2}\alpha \beta
|1m_{s}\right) \left( 11m_{s}m_{l}|2m_{j}\right) Y_{1,m_{l}}\left( \mathbf{n}%
\right) ~,  \label{bm}
\end{equation}%
and are normalized by the condition 
\begin{equation}
\left\langle \mathbf{b}_{m_{j}^{\prime }}^{\ast }\mathbf{b}%
_{m_{j}}\right\rangle =\delta _{m_{j}m_{j}^{\prime }}.  \label{lmnorm}
\end{equation}%
These can be found in the form (See details in Ref. \cite{L12}):%
\begin{align}
\mathbf{b}_{0}& =\sqrt{1/2}\left( -n_{1},-n_{2},2n_{3}\right) ,\mathbf{b}%
_{1}=-\sqrt{3/4}\left( n_{3},in_{3},n_{1}+in_{2}\right) ,  \notag \\
\mathbf{b}_{2}& =\sqrt{3/4}\left( n_{1}+in_{2},in_{1}-n_{2},0\right) ,%
\mathbf{b}_{-m_{j}}=\left( -\right) ^{m_{j}}\mathbf{b}_{m_{j}}^{\ast }.
\label{b012}
\end{align}

In our approach it is necessary to distinguish the interactions in the
channel of two quasiparticles from the interactions in the particle-hole
channel. Since we are interested in values of quasiparticle momenta near the
Fermi surface, $\mathbf{p}\simeq p_{F}\mathbf{n}$, the momentum transferred
in the collision of two quasiparticles is of the order of $2p_{F}$. In this
case the non-central spin-orbit and tensor interactions are most important
in the superdense neutron matter. The attractive spin-orbit interaction
which dominates in the channel of two quasiparticles is the basic reason of
the neutron pairing \cite{Takatsuka}. The most attractive channel
corresponds to spin, orbital, and total angular momenta $s=1$, $l=1$, and $%
j=2$, respectively, and pairs quasiparticles into the $^{3}$P$_{2}$ states
with $m_{j}=0,\pm 1,\pm 2$. The substantially smaller tensor interactions
lift the strong paramagnetic degeneracy inherent in pure $^{3}$P$_{2}$
pairing and mix states of different magnetic quantum numbers. It is well
known that the tensor interactions generate also a small $^{3}$F$_{2}$
admixture to the ground state and markedly modify the $^{3}$P$_{2}$ energy
gap \cite{Khodel,Clark,0203046,SF}.

The purpose of our study is however not the ground state but the linear
response of the superfluid system onto external neutral weak currents. In
this work we assume that the ground state of the superfluid neutron system
and the magnitude of the energy gap at the Fermi surface are the known
external parameters. Neutrino emissivity of the neutron system with a mixed $%
^{3}$P$_{2}$-$^{3}$F$_{2}$ superfluid condensate was examined in Ref. \cite%
{L11c}. According to this work, incorporating of the small admixture of the $%
^{3}$F$_{2}$ state requires of sophisticated calculations but does not
affect noticeably the excitation spectra and the neutrino emissivity through
neutral weak currents. Accordingly, throughout this paper, we neglect tensor
forces. The pairing interaction, in the most attractive channel, can then be
written as \cite{Tamagaki} \ 
\begin{equation}
\varrho \Gamma _{\alpha \beta ,\gamma \delta }\left( \mathbf{p,p}^{\prime
}\right) =V\left( p,p^{\prime }\right) \sum_{m_{j}}\left( \mathbf{b}_{m_{j}}(%
\mathbf{n})\hat{\bm{\sigma}}\hat{g}\right) _{\alpha \beta }\left( \hat{g}%
\hat{\bm{\sigma}}\mathbf{b}_{m_{j}}^{\ast }(\mathbf{n}^{\prime })\right)
_{\gamma \delta }~,  \label{ppint}
\end{equation}%
where $\varrho =p_{F}M^{\ast }/\pi ^{2}$ is the density of states near the
Fermi surface; and $V\left( p,p^{\prime }\right) $ is the interaction
amplitude.

Consider now the interactions in the particle-hole channel. In our analysis,
we shall use the fact that the Fermi-liquid interactions do not interfere
with the pairing phenomenon if approximate hole-particle symmetry is
maintained in the system, i.e., the Fermi-liquid interactions remain
unchanged upon pairing. Near the Fermi surface, the Fermi-liquid effects are
reduced to the standard particle-hole interactions:%
\begin{equation}
\varrho \mathfrak{F}_{\alpha \gamma ,\beta \delta }\left( \mathbf{nn}%
^{\prime }\right) =\mathfrak{f}\left( \mathbf{nn}^{\prime }\right) \delta
_{\alpha \beta }\delta _{\gamma \delta }+\mathfrak{g}\left( \mathbf{nn}%
^{\prime }\right) \bm{\sigma}_{\alpha \beta }\bm{\sigma}_{\gamma \delta }~.
\label{ph}
\end{equation}

For generality, Eq. (\ref{ph}) should be supplemented with contributions
from spin-orbit and tensor interactions. However, in uniform media, the
momentum of particle-hole type excitations equals the transferred momentum $%
q $. We are interested in the medium response at the time-like momentum
transfer, $q\leq \omega \sim \Delta $. In this case the contribution from
the noncentral interactions, which depend on the transferred momentum \cite%
{SF,Migdal}, would be proportional to some power of $q/p_{F}\ll 1$ and
vanish in the limit, $q\rightarrow 0$, which we are interested in (see
below).

\section{Neutrino energy losses}

The emission of neutrino pairs is kinematically possible thanks to the
existence of a superfluid energy gap, which admits the quasiparticle
transitions with time-like momentum transfer $k=\left( \omega ,\mathbf{q}%
\right) $, as required by the final neutrino pair: $k=k_{1}+k_{2}$. We
consider the standard model of weak interactions through neutral weak
currents. After integration over the phase space of escaping neutrinos and
antineutrinos the total energy which is emitted into neutrino pairs per unit
volume and time is given by the following formula (see details, e.g., in
Ref. \cite{L01}): 
\begin{equation}
\epsilon =-\frac{G_{F}^{2}\mathcal{N}}{192\pi ^{5}}\int_{0}^{\infty }d\omega
\int d^{3}q\frac{\omega \Theta \left( \omega -q\right) }{\exp \left( \frac{%
\omega }{T}\right) -1}\operatorname{Im}\Pi _{\mathsf{weak}}^{\mu \nu }\left( \omega ,%
\mathbf{q}\right) \left( k_{\mu }k_{\nu }-k^{2}g_{\mu \nu }\right) ~,
\label{QQQ}
\end{equation}%
where $\Theta \left( x\right) $ is the Heaviside step-function; $%
{\mu}%
,\nu =0,1,2,3$ are Dirac indices; $\mathcal{N}=3$ is the number of neutrino
flavors; $G_{F}$ is the Fermi coupling constant, and $\Pi _{\mathsf{weak}%
}^{\mu \nu }$ is the retarded weak polarization tensor of the medium.

In general, the weak polarization tensor of the medium is a sum of the
vector-vector, axial-axial, and mixed terms. The mixed vector-axial
polarization has to be an antisymmetric tensor, and its contraction in Eq. (%
\ref{QQQ}) with the symmetric tensor $k_{\mu }k_{\nu }-k^{2}g_{\mu \nu }$
vanishes. Thus only the pure-vector and pure-axial polarizations should be
taken into account. We then obtain $\Pi _{\mathsf{weak}}^{\mu \nu }=C_{%
\mathsf{V}}^{2}\Pi _{\mathsf{V}}^{\mu \nu }+C_{\mathsf{A}}^{2}\Pi _{\mathsf{A%
}}^{\mu \nu }$, where $C_{\mathsf{V}}$ and $C_{\mathsf{A}}$ are vector and
axial-vector weak coupling constants of a neutron, respectively.

The Fermi velocity is small in the nonrelativistic system, $V_{F}\ll 1$, and
we can study the neutrino energy losses in the lowest order over this small
parameter. We are interested in the time-like domain of the transferred
energy and momentum, $q<\omega ,$ and $\omega \gtrsim \Delta $, in
accordance with the total energy and momentum of escaping neutrino pairs.
Since the transferred space momentum comes in the polarization functions of
the medium in a combination $\mathbf{qV}_{F}\ll \omega ,\Delta $, one can
evaluate the polarization functions in the limit $\mathbf{q}=0$.
Conservation of the vector current requires $\Pi _{\mathsf{V}}^{\mu \nu
}\left( \omega >0,\mathbf{q}=0\right) =0$. This relation reflects the fact
that the neutrino-pair emission through the vector channel of neutral weak
currents is strongly suppressed in nonrelativistic systems \cite{LP06}.
Therefore we focus on the axial channel of the weak interactions, assuming $%
\Pi _{\mathsf{weak}}^{\mu \nu }\left( \omega ,\mathbf{q}=0\right) \simeq C_{%
\mathsf{A}}^{2}\Pi _{\mathsf{A}}^{\mu \nu }\left( \omega ,\mathbf{q}%
=0\right) $. Further simplification is possible due to the fact that, in the
lowest (zero) order over the particle velocity, only the space component of
the axial-vector vertex survives. This allows to write 
\begin{equation}
\Pi _{\mathsf{weak}}^{\mu \nu }\left( \omega ,\mathbf{q}=0\right) \simeq C_{%
\mathsf{A}}^{2}\delta ^{\mu i}\delta ^{\nu j}\Pi _{\mathsf{A}}^{ij}\left(
\omega ,\mathbf{q}=0\right)  \label{Piw}
\end{equation}%
with $i,j=1,2,3$.

\begin{figure}[h]
\includegraphics{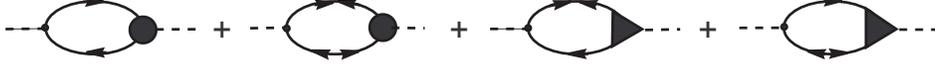}
\caption{Graphs for the polarization tensor. Discarding the residual
particle-hole interactions we show the ordinary vertices points.}
\label{fig1}
\end{figure}

The field interaction with a superfluid should be described with the aid of
four effective three-point vertices. There are two ordinary effective
vertices corresponding to creation of a particle and a hole by the field
that differ by direction of fermion lines and -- two anomalous vertices that
correspond to creation of two particles or two holes. Accordingly, in
graphical representation, the polarization tensor represents a superposition
of loops incorporating the ordinary and anomalous vertices connected by the
ordinary and anomalous Green functions, as depicted in Fig. \ref{fig1}. 
We use the adopted graphical notation for the ordinary and anomalous
propagators, $\hat{G}=\parbox{1cm}{\includegraphics[width=1cm]{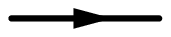}}$, $%
\hat{G}^{-}(p)=\parbox{1cm}{\includegraphics[width=1cm,angle=180]{Gn.eps}}$, 
$\hat{F}^{(1)}=\parbox{1cm}{\includegraphics[width=1cm]{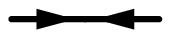}}$\thinspace
, and $\hat{F}^{(2)}=\parbox{1cm}{\includegraphics[width=1cm]{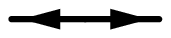}}$%
\thinspace . The analytic form of the quasiparticle propagators in the
momentum representation can be written as 
\begin{equation}
\hat{G}\left( p_{\kappa },\mathbf{p}\right) =G\left( p_{\kappa },\mathbf{p}%
\right) \hat{1},~\ \ \hat{F}^{\left( 1\right) }=\hat{F}\left( p_{\kappa },%
\mathbf{p}\right) =F\left( p_{\kappa },\mathbf{p}\right) \mathbf{\bar{b}}%
\hat{\bm{\sigma}}\hat{g},  \label{eqpr}
\end{equation}%
\begin{equation}
\hat{F}^{\left( 2\right) }=\hat{F}^{\dagger }\left( -p_{\kappa },-\mathbf{p}%
\right) =\hat{g}\hat{\bm{\sigma}}\mathbf{\bar{b}}F\left( -p_{\kappa },-%
\mathbf{p}\right) .  \label{F2}
\end{equation}%
We define the scalar Green functions%
\begin{equation}
G\left( p_{\kappa },\mathbf{p}\right) =\frac{-ip_{\kappa }-\varepsilon _{p}}{%
p_{\kappa }^{2}+E_{\mathbf{p}}^{2}},~\ \ F\left( p_{\kappa },\mathbf{p}%
\right) =F\left( -p_{\kappa },-\mathbf{p}\right) =\frac{\Delta }{p_{\kappa
}^{2}+E_{\mathbf{p}}^{2}}.  \label{GF}
\end{equation}%
where $\mathbf{p}$ is the quasiparticle momentum, and $p_{\kappa }=\left(
2\kappa +1\right) \pi T$ with $\kappa =0,\pm 1,\pm 2,...$ is the fermionic
Matsubara frequency which depends on the temperature $T$. The quasiparticle
energy is given by%
\begin{equation}
E_{\mathbf{p}}=\sqrt{\varepsilon _{p}^{2}+\Delta ^{2}\bar{b}^{2}\left( 
\mathbf{n}\right) }  \label{Eqp}
\end{equation}%
with 
\begin{equation}
\varepsilon _{p}\simeq \upsilon _{F}\left( p-p_{F}\right) .  \label{eps}
\end{equation}

\section{Effective vertices and polarization functions}

\begin{figure}[h]
\includegraphics{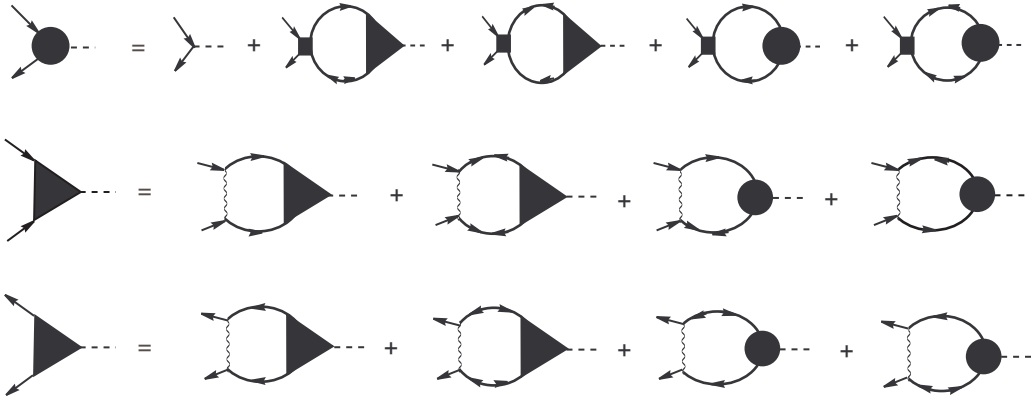}
\caption{Dyson equations for full ordinary and anomalous vertices. The
particle-hole interaction is shown by the shaded rectangle. Wavy lines
represent the pairing interaction.}
\label{fig2}
\end{figure}

The anomalous effective vertices, which we denote as $\mathbf{\hat{T}}%
^{\left( 1\right) }\left( \mathbf{n},\omega \right) $ and $\mathbf{\hat{T}}%
^{\left( 2\right) }\left( \mathbf{n},\omega \right) $, are given by infinite
sums of the diagrams, taking into account the pairing interaction in the
ladder approximation \cite{Nambu}. The ordinary effective vertices, $\hat{%
\bm{\tau}}(\mathbf{n},\omega )\mathbf{~,~}\hat{\bm{\tau}}^{-}\left( \mathbf{n%
},\omega \right) =\hat{\bm{\tau}}^{T}(-\mathbf{n},\omega )$, incorporating
the particle-hole interactions can be evaluated in the random-phase
approximation \cite{Larkin}.
This can be expressed by the set of Dyson equations symbolically depicted by
graphs in Fig. \ref{fig2}, where the particle-hole interactions (\ref{ph})
are shown by the shaded rectangle. Wavy lines represent the pairing
interaction (\ref{ppint}). The first diagram on the right-hand side of the
first line is the three-point vertex of a free particle. In the
nonrelativistic case, the bare axial-vector vertex is given by the spin
matrices $\hat{\bm{\sigma}}$.

The analytic form of these equations, which are to be solved simultaneously
with the gap equations, was repeatedly considered before. The equations can
be substantially simplified due to the fact that we are interested in the
processes near the Fermi surface. After a series of algebraic manipulations
(see details in Ref. \cite{L10c}) the solution valid for $\mathbf{q}=0$ can
be obtained in the form:%
\begin{equation}
\mathbf{\hat{T}}^{\left( 1\right) }=\sum_{m_{j}}\mathbf{B}_{m_{j}}\left(
\omega _{\eta }\right) \left( \bm{\hat{\sigma}}\mathbf{b}_{m_{j}}\right) 
\hat{g},  \label{T1A}
\end{equation}%
\begin{equation}
\mathbf{\hat{T}}^{\left( 2\right) }=\sum_{m_{j}}\mathbf{B}_{m_{j}}\left(
\omega _{\eta }\right) \hat{g}\left( \bm{\hat{\sigma}}\mathbf{b}%
_{m_{j}}\right) ,  \label{T2A}
\end{equation}%
and%
\begin{equation}
\hat{\bm{\tau}}=\phi \left( \mathbf{n},\omega _{\eta }\right) \hat{%
\bm{\sigma}},  \label{tau}
\end{equation}%
where $\omega _{\eta }=2\pi \eta $ with $\eta =0,\pm 1,\pm 2,...$ is the
bosonic Matsubara frequency, and $\phi \left( \mathbf{n},\omega _{\eta
}\right) =\phi \left( -\mathbf{n},\omega _{\eta }\right) $ and $\mathbf{B}%
_{m_{j}}\left( \omega _{\eta }\right) $ are to satisfy the equations (we
omit the $\omega _{\eta }$ dependence of the functions):%
\begin{equation}
\mathbf{B}_{m_{j}}=-\frac{i\omega _{\eta }}{\Delta }\frac{1}{\chi _{m_{j}}}%
\left\langle i\left( \mathbf{b}_{m_{j}}^{\ast }\mathbf{\times \bar{b}}%
\right) \mathcal{I}_{0}\phi \right\rangle ,  \label{RP}
\end{equation}%
\begin{equation}
\phi =1+\frac{1}{3}\int \frac{d\mathbf{n}^{\prime }}{4\pi }\mathfrak{g}%
\left( \mathbf{nn}^{\prime }\right) \left[ \sum_{m_{j}}\frac{i\omega _{\eta }%
}{\Delta }\mathcal{I}_{0}i\left( \mathbf{b}_{m_{j}}\mathbf{\times \bar{b}}%
\right) \cdot \mathbf{B}_{m_{j}}-4\bar{b}^{2}\mathcal{I}_{0}\phi \right] _{%
\mathbf{n}^{\prime }}~  \label{fi}
\end{equation}%
with%
\begin{equation}
\chi _{m_{j}}\equiv \left\langle \left( \mathbf{b}_{m_{j}}^{\ast }\mathbf{b}%
_{m}-\bar{b}^{2}\right) A\right\rangle -\frac{\omega _{n}^{2}}{2\Delta ^{2}}%
\left\langle \left( \mathbf{b}_{m_{j}}^{\ast }\mathbf{b}_{m}\right) \mathcal{%
I}_{0}\right\rangle -2\left\langle \left( \mathbf{b}_{m_{j}}^{\ast }\mathbf{%
\bar{b}}\right) \left( \mathbf{\bar{b}b}_{m_{j}}\right) \mathcal{I}%
_{0}\right\rangle .  \label{him}
\end{equation}%
The functions $A\left( \mathbf{n}\right) $ and $\mathcal{I}_{0}\left( \omega
_{\eta }\mathbf{;n}\right) $ are given by 
\begin{equation}
A\left( \mathbf{n}\right) \equiv \int_{-\infty }^{\infty }d\varepsilon
\left( \frac{1}{2E}\tanh \frac{E}{2T}-\frac{1}{2\varepsilon }\tanh \frac{%
\varepsilon }{2T}\right) ,  \label{Aex}
\end{equation}%
and 
\begin{equation}
\mathcal{I}_{0}\left( \omega _{\eta }\mathbf{;n}\right) =\int_{-\infty
}^{\infty }\frac{d\varepsilon }{E}\frac{\Delta ^{2}}{4E^{2}+\omega _{\eta
}^{2}}\tanh \frac{E}{2T}~,  \label{FFq0}
\end{equation}%
with 
\begin{equation}
E=\sqrt{\varepsilon ^{2}+\Delta ^{2}\bar{b}^{2}\left( \mathbf{n}\right) }~\ .
\label{E}
\end{equation}

To solve the set of Eqs. (\ref{RP}), (\ref{fi}) we expand $\phi \left( 
\mathbf{n}\right) $ in spherical harmonics%
\begin{equation}
\phi \left( \mathbf{n,}\omega _{n}\right) =\sum_{lm}\phi _{l,m}\left( \omega
_{n}\right) Y_{l,m}\left( \theta ,\varphi \right) ,  \label{fiY}
\end{equation}%
where $\theta $ and $\varphi $ are, respectively, the polar and azimuthal
angles of the vector $\mathbf{n}$, and write%
\begin{equation}
\mathfrak{g}\left( \mathbf{nn}^{\prime }\right) =\sum_{l=0}^{\infty }\frac{%
4\pi \mathfrak{g}_{l}}{2l+1}\sum_{\mu }Y_{l,\mu }\left( \theta ,\varphi
\right) Y_{l,\mu }^{\ast }\left( \theta ^{\prime },\varphi ^{\prime }\right)
,  \label{gY}
\end{equation}

Inserting (\ref{fiY}) and (\ref{gY}) into Eqs. (\ref{RP}) and (\ref{fi}) we
obtain 
\begin{equation}
\mathbf{B}_{m_{j}}=-\frac{i\omega _{\eta }}{\Delta }\frac{1}{\chi _{m_{j}}}%
\sum_{lm}\phi _{l,m}\left\langle \mathcal{I}_{0}i\left( \mathbf{b}%
_{m_{j}}^{\ast }\mathbf{\times \bar{b}}\right) Y_{l,m}\right\rangle
\label{Bmj}
\end{equation}%
and%
\begin{equation}
\phi _{l,m}=\sqrt{4\pi }\delta _{l,0}\delta _{m,0}+\frac{\mathfrak{g}_{l}}{%
2l+1}\frac{4}{3}\sum_{l^{\prime }m^{\prime }}\left( \frac{\omega _{\eta }^{2}%
}{4\Delta ^{2}}\lambda _{l,m;l^{\prime },m^{\prime }}-\eta _{l,m;l^{\prime
},m^{\prime }}\right) \phi _{l^{\prime },m^{\prime }},  \label{film}
\end{equation}%
where we defined dimensionless quantities%
\begin{equation}
\lambda _{l,m;l^{\prime },m^{\prime }}\equiv 4\pi \sum_{m_{j}}\frac{1}{\chi
_{m_{j}}}\left\langle Y_{l,m}^{\ast }\mathcal{I}_{0}i\left( \mathbf{b}%
_{m_{j}}\mathbf{\times \bar{b}}\right) \right\rangle \cdot \left\langle
Y_{l^{\prime },m^{\prime }}\mathcal{I}_{0}i\left( \mathbf{b}_{m_{j}}^{\ast }%
\mathbf{\times \bar{b}}\right) \right\rangle ,  \label{lam}
\end{equation}%
and%
\begin{equation}
\eta _{l,m;l^{\prime },m^{\prime }}\equiv 4\pi \left\langle Y_{l,m}^{\ast }%
\bar{b}^{2}\mathcal{I}_{0}Y_{l^{\prime },m^{\prime }}\right\rangle .
\label{eta}
\end{equation}%
We will focus on the condensation into the state $^{3}P_{2}$ with $m_{j}=0$
which is conventionally considered as the preferable one in the bulk matter
of neutron stars. In this case%
\begin{equation}
\mathbf{\bar{b}}\left( \mathbf{n}\right) =\mathbf{b}_{0}\left( \mathbf{n}%
\right) ~,~\ \ \bar{b}^{2}=\frac{1}{2}\left( 1+3n_{3}^{2}\right)  \label{b0}
\end{equation}%
Since the function $\mathcal{I}_{0}\left( \omega _{n},\bar{b}^{2}\right) $
is axial symmetric the integration over the azimuthal angle can be done in
Eqs. (\ref{lam}) and (\ref{eta}). By performing subsequent summation over $%
m_{j}=0,\pm 1,\pm 2$ we find that only $m_{j}=\pm 1$ contribute, and $\chi
_{-1}=\chi _{1}\equiv \chi $. In this way we obtain (the $\omega _{\eta }$
dependence is omitted):%
\begin{equation}
\lambda _{l,m;l^{\prime },m^{\prime }}=-\delta _{m,0}\delta _{m^{\prime },0}%
\frac{3}{2}\frac{\beta _{l}\beta _{l^{\prime }}}{\chi }  \label{laml}
\end{equation}%
with%
\begin{equation}
\beta _{l}\left( \omega _{\eta }\right) \equiv \left\langle \bar{b}%
^{2}\left( n_{3}^{2}\right) \mathcal{I}_{0}\left( \omega _{\eta
},n_{3}^{2}\right) P_{l}\left( n_{3}\right) \right\rangle ,  \label{beta}
\end{equation}%
where $P_{l}\left( x\right) $ are the Legendre polynomials, and%
\begin{equation}
\eta _{l,m;l^{\prime },m^{\prime }}=\delta _{m,m^{\prime }}\eta
_{l,m;l^{\prime },m}.  \label{etal}
\end{equation}

In these notation we obtain the Eq. (\ref{film}) in the form%
\begin{equation}
\phi _{l,m}=\sqrt{4\pi }\delta _{l,0}\delta _{m,0}-\frac{\mathfrak{g}_{l}}{%
2l+1}\frac{4}{3}\sum_{l^{\prime }}\left( \delta _{m,0}\frac{3}{2}\frac{%
\omega _{\eta }^{2}}{4\Delta ^{2}}\frac{\beta _{l}\beta _{l^{\prime }}}{\chi 
}\ +\eta _{l,m;l^{\prime },m}\right) \phi _{l^{\prime },m},  \label{fieq}
\end{equation}%
For $m\neq 0$, this equation has only trivial solution $\phi _{l,m\neq 0}=0$%
. For $\phi _{l}\equiv \phi _{l,0}$ Eq. (\ref{fieq}) takes the form 
\begin{equation}
\phi _{l}=\sqrt{4\pi }\delta _{l,0}-\frac{\mathfrak{g}_{l}}{2l+1}\frac{4}{3}%
\sum_{l^{\prime }}\left( \frac{3}{2}\frac{\omega _{\eta }^{2}}{4\Delta ^{2}}%
\frac{\beta _{l}\beta _{l^{\prime }}}{\chi }+\gamma _{l,l^{\prime }}\right)
\phi _{l^{\prime }}  \label{fil}
\end{equation}%
with $\beta _{l}$, as given in Eq. (\ref{beta}) and 
\begin{equation}
\gamma _{l,l^{\prime }}\left( \omega _{\eta }\right) \equiv \left\langle
P_{l}\left( n_{3}\right) \bar{b}^{2}\left( n_{3}^{2}\right) \mathcal{I}%
_{0}\left( \omega _{\eta },n_{3}^{2}\right) P_{l^{\prime }}\left(
n_{3}\right) \right\rangle .  \label{gam}
\end{equation}%
Since the $\phi \left( \mathbf{n},\omega \right) =\phi \left( -\mathbf{n}%
,\omega \right) $ only even values of $l$ contribute into the expansion (\ref%
{fiY}) which takes the form%
\begin{equation}
\phi \left( \mathbf{n,}\omega _{\eta }\right) =\frac{1}{\sqrt{4\pi }}\sum_{l=%
\mathsf{even}}\phi _{l}\left( \omega _{\eta }\right) P_{l}\left(
n_{3}\right) .  \label{Fi}
\end{equation}

The Fermi-liquid parameters are not well known for a wide range of neutron
densities we consider. Some known data allow to hope, however, that, in Eq. (%
\ref{fil}), the parameters $\mathfrak{g}_{l}/\left( 2l+1\right) $ decrease
rapidly for $l\geq 2$. For example, according to Ref. \cite{SF}, at the
Fermi momentum $p_{F}=1.7$ fm$^{-1}$ one has $\mathfrak{g}_{0}=0.842$ while $%
\frac{1}{5}\mathfrak{g}_{2}=0.043\,8$. Therefore we take the approximation $%
\mathfrak{g}_{l}=0$ for $l\geq 2$ thus obtaining 
\begin{equation}
\phi =\frac{\chi }{\left( 1+\frac{4}{3}\mathfrak{g}_{0}\left\langle \bar{b}%
^{2}\mathcal{I}_{0}\right\rangle \right) \chi +2\mathfrak{g}_{0}\Omega
_{\eta }^{2}\left\langle \bar{b}^{2}\mathcal{I}_{0}\right\rangle ^{2}}.
\label{fi0}
\end{equation}%
Inserting this expression into Eq. (\ref{RP}) gives $\mathbf{B}_{0}=\mathbf{B%
}_{\pm 2}=0,$ 
\begin{equation}
\mathbf{B}_{\pm 1}\left( \omega _{\eta }\right) =-\frac{i\omega _{\eta }}{%
2\Delta }\sqrt{\frac{3}{2}}\frac{\left\langle \bar{b}^{2}\mathcal{I}%
_{0}\right\rangle }{\left( 1+\frac{4}{3}\mathfrak{g}_{0}\left\langle \bar{b}%
^{2}\mathcal{I}_{0}\right\rangle \right) \chi +2\mathfrak{g}_{0}\Omega
_{\eta }^{2}\left\langle \bar{b}^{2}\mathcal{I}_{0}\right\rangle ^{2}}\left( 
\begin{array}{c}
1 \\ 
\mp i \\ 
0%
\end{array}%
\right) ,  \label{B1}
\end{equation}%
with%
\begin{equation}
\chi \equiv \left\langle 2\left[ -\Omega _{\eta }^{2}\mathbf{b}_{1}^{\ast }%
\mathbf{b}_{1}-\left( \mathbf{b}_{1}^{\ast }\mathbf{\bar{b}}\right) \left( 
\mathbf{b}_{1}\mathbf{\bar{b}}\right) \right] \mathcal{I}_{0}+\left( \mathbf{%
b}_{1}^{\ast }\mathbf{b}_{1}-\bar{b}^{2}\right) A\right\rangle \text{.}
\label{hi1}
\end{equation}%
In obtaining Eq. (\ref{B1}) we used the identities $\beta _{0}=\gamma
_{0,0}\equiv \left\langle \bar{b}^{2}\mathcal{I}_{0}\right\rangle $. After
the replacement $i\omega _{n}\rightarrow \omega +i0$ we obtain the analytic
continuation to the retarded vertex for particles. The replacement $i\omega
_{n}\rightarrow \omega -i0$ gives the advanced vertex for holes.

From Eqs. (\ref{T1A}), (\ref{T2A}) and (\ref{tau}) we find%
\begin{equation}
\mathbf{\hat{T}}^{\left( 1\right) }\left( \mathbf{n},\Omega \right)
=-f\left( \Omega ,y\right) \left[ \mathbf{e}^{\ast }\left( \bm{\hat{\sigma}}%
\mathbf{b}_{1}\right) \hat{g}+\mathbf{e}\left( \bm{\hat{\sigma}}\mathbf{b}%
_{-1}\right) \hat{g}\right] ,  \label{T1}
\end{equation}%
\begin{equation}
\mathbf{\hat{T}}^{\left( 2\right) }\left( \mathbf{n},\Omega \right)
=-f^{\ast }\left( \Omega ,y\right) \left[ \mathbf{e}^{\ast }\hat{g}\left( %
\bm{\hat{\sigma}}\mathbf{b}_{1}\right) +\mathbf{e}\hat{g}\left( %
\bm{\hat{\sigma}}\mathbf{b}_{-1}\right) \right]  \label{T2}
\end{equation}%
and%
\begin{equation}
\hat{\bm{\tau}}=\phi \left( \Omega ,y\right) \hat{\bm{\sigma}},
\label{drtau}
\end{equation}%
\begin{equation}
\hat{\bm{\tau}}^{-}=\phi ^{\ast }\left( \Omega ,y\right) \hat{\bm{\sigma}^{T}%
}  \label{tauh}
\end{equation}%
with $\mathbf{e}=\left( 1,i,0\right) $, 
\begin{equation}
\Omega =\frac{\omega }{2\Delta \left( T\right) },~y=\frac{\Delta \left(
T\right) }{T}.  \label{Oy}
\end{equation}%
The functions $f\left( \Omega ,y\right) $ and $\phi \left( \Omega ,y\right) $
are given by 
\begin{equation}
f\left( \Omega ,y\right) \equiv \sqrt{\frac{3}{2}}\frac{\Omega \left\langle 
\bar{b}^{2}\mathcal{I}_{0}\right\rangle }{\left( 1+\frac{4}{3}\mathfrak{g}%
_{0}\left\langle \bar{b}^{2}\mathcal{I}_{0}\right\rangle \right) \chi -2%
\mathfrak{g}_{0}\Omega ^{2}\left\langle \bar{b}^{2}\mathcal{I}%
_{0}\right\rangle ^{2}}  \label{fy}
\end{equation}%
and%
\begin{equation}
\phi \left( \Omega ,y\right) \equiv \frac{\chi }{\left( 1+\frac{4}{3}%
\mathfrak{g}_{0}\left\langle \bar{b}^{2}\mathcal{I}_{0}\right\rangle \right)
\chi -2\mathfrak{g}_{0}\Omega ^{2}\left\langle \bar{b}^{2}\mathcal{I}%
_{0}\right\rangle ^{2}},  \label{fiy}
\end{equation}%
where%
\begin{eqnarray}
\chi \left( \Omega ,y\right) &=&\frac{1}{4}\left[ \int_{0}^{1}dn_{3}\left[
6\Omega ^{2}\allowbreak \left( 1+n_{3}^{2}\right) -~3n_{3}^{2}\left(
1-n_{3}^{2}\right) \right] \mathcal{I}_{0}\left( n_{3}^{2},\Omega ,y\right)
\right.  \notag \\
&&\left. +\int_{0}^{1}dn_{3}\left( 1-3n_{3}^{2}\right) A\left(
n_{3}^{2},y\right) \right] ,  \label{hi}
\end{eqnarray}%
and%
\begin{equation}
\left\langle \bar{b}^{2}\mathcal{I}_{0}\right\rangle =\frac{1}{2}%
\int_{0}^{1}dn_{3}\left( 1+3n_{3}^{2}\right) \mathcal{I}_{0}.  \label{b2i}
\end{equation}%
In obtaining Eq. (\ref{T2}) we used the identity $\mathcal{I}_{0}\left(
\omega -i0\right) =\mathcal{I}_{0}^{\ast }\left( \omega +i0\right) $.

Making use of the effective vertices (\ref{T1})-(\ref{tauh}) and the
propagators (\ref{eqpr})-(\ref{GF}) one can calculate the axial polarization
tensor by the diagrams of Fig. \ref{fig1} (See details in Refs. \cite%
{L10a,L10b}). For $q=0$ we obtain: 
\begin{align}
\Pi _{\mathrm{A}}^{ij}\left( \Omega ,y\right) & =-4\rho \left[ \operatorname{Re}\phi
\left( \Omega ,y\right) \left\langle \left( \delta _{ij}-\frac{\bar{b}_{i}%
\bar{b}_{j}}{\bar{b}^{2}}\right) \bar{b}^{2}\mathcal{I}_{0}\right\rangle
\right.  \notag \\
& \left. -\left( \delta _{ij}-\delta _{i3}\delta _{j3}\right) \sqrt{\frac{3}{%
2}}\Omega \operatorname{Re}f\left( \Omega \right) \left\langle \bar{b}^{2}\mathcal{I}%
_{0}\right\rangle \right] ~.  \label{AKij}
\end{align}%
with $\mathbf{\bar{b}}\left( \mathbf{n}\right) =\sqrt{1/2}\left(
-n_{1},-n_{2},2n_{3}\right) $ and $\bar{b}^{2}=\frac{1}{2}\left(
1+3n_{3}^{2}\right) $, as given by Eq. (\ref{b0}).

The imaginary part of $\Pi _{\mathrm{A}}^{ij}$ that arises owing to the PBF
processes originates from the function $\mathcal{I}_{0}\left( n_{3};\Omega
,y\right) $ at $\Omega >\bar{b}_{\min }=1/\sqrt{2}$. A one more contribution
into the imaginary part of the axial polarization tensor (\ref{AKij}) arises
from the pole of the function $f\left( \Omega ,y\right) $ at $\Omega =\Omega
_{s}<\bar{b}_{\min }$. This contribution describes the neutrino energy
losses caused by the spin wave decays. The PBF and SWD processes operate in
different kinematical domains, so that the imaginary part of the
polarization tensor consists of two clearly distinguishable contributions, $%
\operatorname{Im}\Pi _{\mathrm{A}}^{ij}=\operatorname{Im}\Pi _{\mathrm{PBf}}^{ij}+\operatorname{Im}%
\Pi _{\mathrm{SWD}}^{ij}$, which we now consider.

\section{Neutrino losses}

\subsection{PBF channel}

First we examine the PBF processes occuring at $\omega >2\Delta _{\mathbf{n}%
}\equiv 2\bar{b}\left( \mathbf{n}\right) \Delta $ or, equivalently, at $%
\Omega >\bar{b}_{\min }$. From Eq. (\ref{AKij}) we obtain%
\begin{align}
\operatorname{Im}\Pi _{\mathrm{A}}^{ij}\left( \Omega >\frac{1}{\sqrt{2}}\right) &
=-4\rho \left[ \operatorname{Re}\phi ~\left\langle \left( \delta _{ij}-\frac{\bar{b}%
_{i}\bar{b}_{j}}{\bar{b}^{2}}\right) \bar{b}^{2}\operatorname{Im}\mathcal{I}%
_{0}\right\rangle \right.  \notag \\
& \left. -4\rho \left( \delta _{ij}-\delta _{i3}\delta _{j3}\right) \sqrt{%
\frac{3}{2}}\Omega \operatorname{Re}f~\,\left\langle \bar{b}^{2}\operatorname{Im}\mathcal{I}%
_{0}\right\rangle \right]  \label{ImPiA}
\end{align}%
Inserting this expression into Eqs. (\ref{Piw}) and (\ref{QQQ}) we calculate
the contraction of $\operatorname{Im}\Pi _{\mathsf{weak}}^{\mu \nu }$ with symmetric
tensor $k_{\mu }k_{\nu }-k^{2}g_{\mu \nu }$ to obtain%
\begin{align}
\operatorname{Im}\Pi _{\mathsf{weak}}^{\mu \nu }\left( k_{\mu }k_{\nu }-k^{2}g_{\mu
\nu }\right) & \simeq -4\rho \left( \operatorname{Re}\phi -\sqrt{\frac{3}{2}}\Omega 
\operatorname{Re}f\right)  \notag \\
& \times \left\langle \left( 2\omega ^{2}-2q_{\parallel }^{2}-q_{\perp
}^{2}\right) \bar{b}^{2}\operatorname{Im}\mathcal{I}_{0}\right\rangle ,
\label{contr}
\end{align}%
where we use the local frame with $Oz\parallel \mathbf{\bar{b}}$, and $%
q_{\parallel }$ and $q_{\perp }$ are defined as%
\begin{equation}
q_{\parallel }^{2}=\frac{1}{\bar{b}^{2}}\left( \mathbf{q\bar{b}}\right)
^{2}~,~q_{\perp }^{2}=q^{2}-q_{\parallel }^{2}~.  \label{qsq}
\end{equation}%
The functions $\mathcal{I}_{0}$, $\phi $ and $f$ are independent of the
space momentum of the neutrino pair. Therefore the integral over $d^{3}q$ in
Eq. (\ref{QQQ}) can be performed, and we obtain the neutrino emissivity in
the form:%
\begin{equation}
\epsilon =\frac{32}{15\pi ^{6}}C_{\mathsf{A}}^{2}G_{F}^{2}p_{F}M^{\ast}\mathcal{N}T^{7}\int_{0}^{\infty }d\Omega \frac{y^{7}\Omega ^{6}}{\exp \left(
2y\Omega \right) -1}\left( \operatorname{Re}\phi -\sqrt{\frac{3}{2}}\Omega \operatorname{Re}%
f\right) \left\langle \bar{b}^{2}\operatorname{Im}\mathcal{I}_{0}\right\rangle .
\label{eps1}
\end{equation}

Imaginary part of the analytic continuation of the function (\ref{FFq0}) is
given by 
\begin{equation}
\operatorname{Im}\mathcal{I}_{0}\left( \omega +i0\right) =-\operatorname{sign}\left( \Omega
\right) \frac{\pi ~\Theta \left( \Omega ^{2}-\bar{b}^{2}\right) }{4\Omega 
\sqrt{\Omega ^{2}-\bar{b}^{2}}}\tanh \left( \frac{y\Omega }{2}\right) .
\label{imI0}
\end{equation}%
With $\bar{b}^{2}$ given in Eq. (\ref{b0}) one finds for $\Omega ^{2}>1/2$: 
\begin{align}
\left\langle \bar{b}^{2}\operatorname{Im}\mathcal{I}_{0}\right\rangle & =\frac{\pi
^{2}\sqrt{6}}{96}\frac{1}{\Omega }\left( 1+2\Omega ^{2}\right) \tanh \left( 
\frac{y\Omega }{2}\right) \Theta \left( \Omega ^{2}-\frac{1}{2}\right)
\Theta \left( 2-\Omega ^{2}\right)  \notag \\
& +\frac{\pi \sqrt{6}}{48}\frac{1}{\Omega }\left[ \left( 2\Omega
^{2}+1\right) \arcsin \sqrt{\frac{3}{2\Omega ^{2}-1}}-\sqrt{6\Omega ^{2}-12}%
\right]  \notag \\
& \times \tanh \left( \frac{y\Omega }{2}\right) \Theta \left( \Omega
^{2}-2\right)  \label{b2Im}
\end{align}%
The integration over $d\Omega $ is now parted into two intervals: 
\begin{align}
\epsilon & =\epsilon _{0}\tau ^{7}y^{7}\left[ 4\int_{1/\sqrt{2}}^{\sqrt{2}%
}d\Omega \frac{\left( 1+2\Omega ^{2}\right) \Omega ^{5}}{\left( e^{y\Omega
}+1\right) ^{2}}\left( \operatorname{Re}\phi -\sqrt{\frac{3}{2}}\Omega ^{2}\operatorname{Re}%
f\right) \right.  \notag \\
& +\frac{8}{\pi }\int_{\sqrt{2}}^{\infty }d\Omega \frac{\Omega ^{5}}{\left(
e^{y\Omega }+1\right) ^{2}}\left( \operatorname{Re}\phi -\sqrt{\frac{3}{2}}\Omega
^{2}\operatorname{Im}f\right)  \notag \\
& \left. \times \left( \left( 2\Omega ^{2}+1\right) \arcsin \frac{\sqrt{3}}{%
\sqrt{2\Omega ^{2}-1}}-\sqrt{6\Omega ^{2}-12}\right) \right] ,  \label{eps2}
\end{align}%
where we denote $\tau =T/T_{c}$, and $y=y\left( \tau \right) $;%
\begin{eqnarray}
\epsilon _{0} &\equiv &\frac{\sqrt{6}}{180\pi ^{4}}C_{\mathsf{A}%
}^{2}G_{F}^{2}p_{F}M^{\ast }\mathcal{N} T_{c}^{7}  \notag \\
&=&1.\,\allowbreak 88\,\times 10^{20}~\left( \frac{M^{\ast }}{M}\right)
\left( \frac{p_{F}}{Mc}\right) T_{9c}^{7}\mathcal{N}_{\nu }C_{\mathrm{A}%
}^{2}~\frac{\mathsf{erg}}{\mathsf{cm}^{3}\mathsf{s}},  \label{eps0}
\end{eqnarray}%
$M$ and $M^{\ast }$ are the effective and bare nucleon masses, respectively; 
$T_{9c}\equiv T_{c}/10^{9}\mathsf{K}$.

Equation (\ref{eps2}) with $f\left( \Omega ,y\right) $ and $\phi \left(
\Omega ,y\right) $ as given in Eqs. (\ref{fy}) and (\ref{fiy}) improves
previous results obtained in Ref. \cite{L10a}, where the vertex function was
evaluated in the BCS and average-angle approximation yielding 
\begin{equation}
f_{\mathsf{av}}\left( \Omega \right) \simeq \frac{1}{2}\sqrt{\frac{3}{2}}%
\frac{1}{\Omega },~\phi _{\mathsf{BCS}}=1\text{.}  \label{fav}
\end{equation}%
Therefore before proceeding to the detailed analysis of the neutrino losses
caused by the PBF processes, we examine the obtained equations for the above
approximations.

Replacing $f\left( \Omega \right) \rightarrow f_{\mathsf{av}}\left( \Omega
\right) $ in Eq. (\ref{eps2}) leads to the following result

\begin{align}
\epsilon _{\mathsf{av}}\left( \tau \right) & =\epsilon _{0}\tau ^{7}y^{7}%
\left[ \int_{1/\sqrt{2}}^{\sqrt{2}}d\Omega \frac{\left( 1+2\Omega
^{2}\right) \Omega ^{5}}{\left( e^{y\Omega }+1\right) ^{2}}\right.  \notag \\
& \left. +\frac{2}{\pi }\int_{\sqrt{2}}^{\infty }d\Omega \frac{\Omega ^{5}}{%
\left( e^{y\Omega }+1\right) ^{2}}\left( \left( 2\Omega ^{2}+1\right)
\arcsin \frac{\sqrt{3}}{\sqrt{2\Omega ^{2}-1}}-\sqrt{6\Omega ^{2}-12}\right) %
\right] .  \label{epsav}
\end{align}%
The neutrino emissivity, as found in Ref. \cite{L10a} in the average-angle
approximation, is written in the form of the two-fold integral over the
Fermi surface and over the energy of escaping neutrino pairs. 
\begin{figure}[h]
\includegraphics{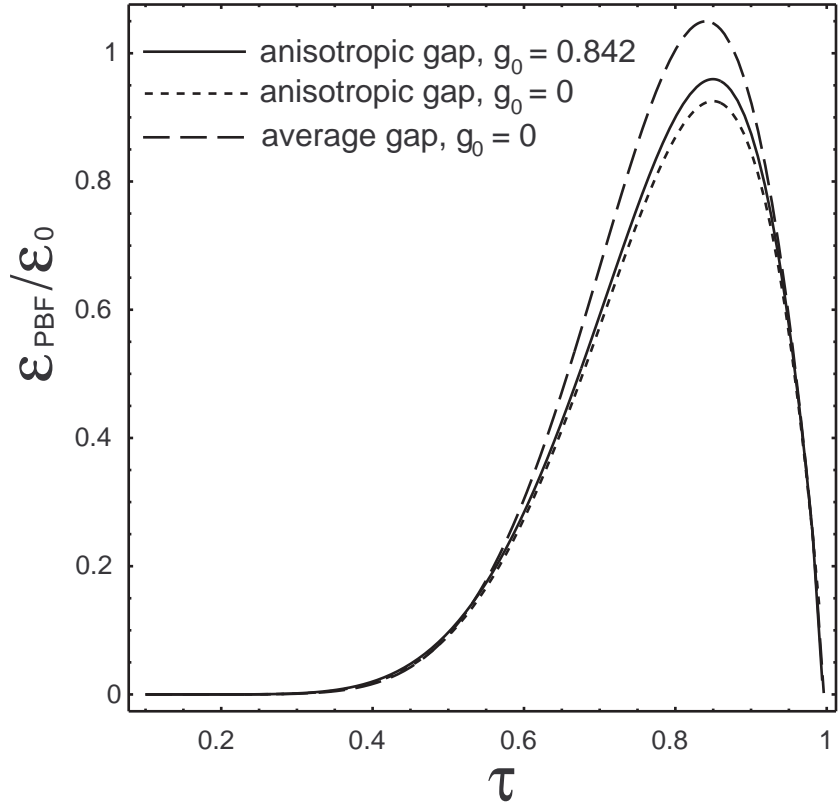}
\caption{PBF neutrino emissivity in units $\protect\epsilon _{0}$ versus the
dimensionless temperature $\protect\tau =T/T_{c}$ in various approximations
marked in the plot.}
\label{fig3}
\end{figure}
The integration over the Fermi surface in that expression can be performed
analytically. In the Appendix, we demonstrate that such the integration
results in Eq. (\ref{epsav})

For numerical evaluation of the neutrino losses, with making use of Eqs. (%
\ref{eps2}) and (\ref{epsav}) it is necessary to know the function $y\left(
\tau \right) $ which in general is to be found with the aid of the gap
equation. For the case $m_{j}=0$ the function is well investigated. We can
adjust for example the simple fit to $y\left( \tau \right) =\sqrt{2\left(
1-\tau \right) }\left( 0.7893+1.188/\tau \right) $, as suggested in Ref. 
\cite{YKL}.

In Fig. \ref{fig3} we show the neutrino energy losses computed according to
the exact Eqs. (\ref{eps2}, \ref{fy}, \ref{fiy}) in comparison with the
energy losses in the average-angle approximation, as given by Eq. (\ref%
{epsav}). The gap anisotropy leads to quenching of the PBF neutrino
emissivity on about 15\%, while the exchange particle-hole interactions with 
$\mathfrak{g}_{0}=0.842$ slightly increase the neutrino energy losses. The
neutrino emissivity which incorporates both the effects is less than the
result obtained in the average-angle approximation on approximately 10 \%.

\subsection{SWD channel}

The function $\mathcal{I}_{0}\left( n_{3};\Omega ,y\right) $ is real at $%
\Omega <\bar{b}_{\min }$. From Eq. (\ref{ImPiA}), (\ref{fy}) and (\ref{fiy})
we find in this domain%
\begin{align}
\Pi _{\mathsf{weak}}^{\mu \nu }\left( k_{\mu }k_{\nu }-k^{2}g_{\mu \nu
}\right) & \simeq -4\rho \left( \chi -\frac{3}{2}\Omega ^{2}\left\langle 
\bar{b}^{2}\mathcal{I}_{0}\right\rangle \right)  \notag \\
& \times \frac{\left\langle \left( 2\omega ^{2}-2q_{\parallel }^{2}-q_{\perp
}^{2}\right) \bar{b}^{2}\mathcal{I}_{0}\right\rangle }{\left( 1+\frac{4}{3}%
\mathfrak{g}_{0}\left\langle \bar{b}^{2}\mathcal{I}_{0}\right\rangle \right)
\chi -2\mathfrak{g}_{0}\Omega ^{2}\left\langle \bar{b}^{2}\mathcal{I}%
_{0}\right\rangle ^{2}},  \label{Pmn}
\end{align}%
This function has a pole owing to existence of eigen oscillations of the
condensate at the frequency $\omega _{s}$ satisfying the condition 
\begin{equation}
\left( 1+\frac{4}{3}\mathfrak{g}_{0}\left\langle \bar{b}^{2}\mathcal{I}%
_{0}\right\rangle \right) \chi -2\mathfrak{g}_{0}\Omega _{s}^{2}\left\langle 
\bar{b}^{2}\mathcal{I}_{0}\right\rangle ^{2}=0,  \label{omegas}
\end{equation}%
where all the functions are to be taken at $\Omega =\Omega _{s}$, defined as 
\begin{equation}
\Omega _{s}=\frac{\omega _{s}}{2\Delta \left( T\right) }.  \label{Omegas}
\end{equation}%
As discussed in Ref. \cite{L12}, the corresponding nonunitary oscillations
look like the "normal-flapping" mode in $^{3}$He-A \cite{W}.

Solution to Eq. (\ref{omegas}) can by found by assuming that the oscillation
frequency of this wave is small, $\Omega _{s}^{2}\ll 1$. This allows to
neglect $\omega ^{2}$ in the integrand of Eq. (\ref{FFq0}) and write%
\begin{equation}
\mathcal{I}_{0}\left( n_{3};\omega _{s}\right) \simeq \tilde{I}_{0}\left(
n_{3}\right) \equiv \frac{\Delta ^{2}}{4}\int_{-\infty }^{\infty }\frac{%
d\varepsilon }{E^{3}}\tanh \frac{E}{2T}.  \label{It}
\end{equation}%
After this simplification\ the analytic solution to Eq. (\ref{omegas}) can
be written as:%
\begin{equation}
\Omega _{s}^{2}=\frac{\int_{0}^{1}dn_{3}\left[ 3n_{3}^{2}\left(
1-n_{3}^{2}\right) \tilde{I}_{0}-\left( 1-3n_{3}^{2}\right) A\right] }{%
6\int_{0}^{1}dn_{3}\left( 1+n_{3}^{2}\right) \tilde{I}_{0}-8\mathfrak{g}%
_{0}\left\langle \bar{b}^{2}\tilde{I}_{0}\right\rangle ^{2}\left( 1+\frac{4}{%
3}\mathfrak{g}_{0}\left\langle \bar{b}^{2}\tilde{I}_{0}\right\rangle \right)
^{-1}},  \label{w1}
\end{equation}%
where the function $A$ is given by Eq. (\ref{Aex}), and%
\begin{equation}
\left\langle \bar{b}^{2}\tilde{I}_{0}\right\rangle =\frac{1}{2}%
\int_{0}^{1}dn_{3}\left( 1+3n_{3}^{2}\right) \tilde{I}_{0}.  \label{b2it}
\end{equation}%

\begin{figure}[h]
\includegraphics{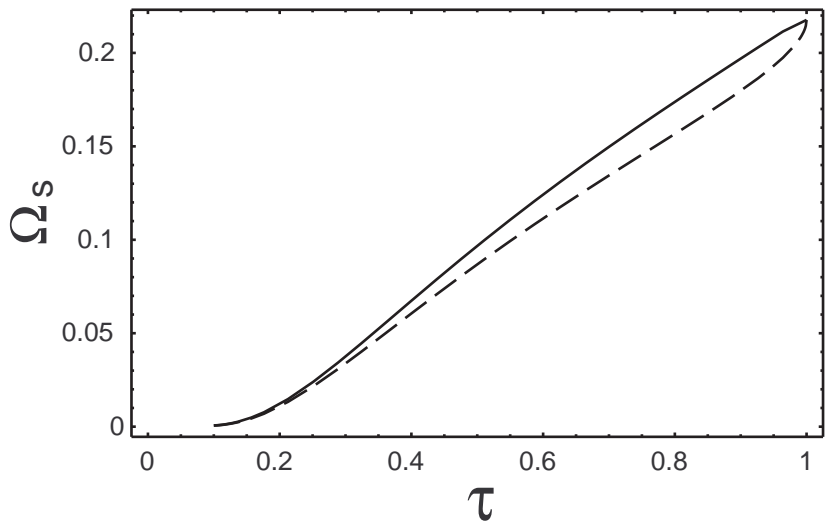}
\caption{The frequency $\protect\omega _{s}\left( \protect\tau \right) $ of
the spin wave at $q=0$ in units of $2\Delta \left( \protect\tau \right) $
versus reduced temperature $\protect\tau =T/T_{c}$. Solid curve is
calculated for $\mathfrak{g}_{0}=0.842$. Long-dashed curve corresponds to $%
\mathfrak{g}_{0}=0$.}
\label{fig4}
\end{figure}

In Ref. \cite{L10c} the right-hand side of Eq. (\ref{w1}) was evaluated for $%
\mathfrak{g}_{0}=0$ in the average-angle approximation assuming that the
anisotropic gap in the quasiparticle energy is replaced by its average-angle
magnitude, $\bar{b}^{2}\Delta ^{2}\left( T\right) \rightarrow \Delta
^{2}\left( T\right) $. Such approach results in $\Omega _{s}^{2}\rightarrow
1/20$ or, equivalently, $\omega _{s}^{\mathsf{av}}\simeq \Delta \left(
T\right) /\sqrt{5}$ with a simple temperature dependence of the excitation
energy only through the gap amplitude. However, the more accurate
calculation \cite{L12} has shown that, for the eigen modes, the
average-angle approximation is valid only in the limit $T\rightarrow T_{c}$%
.\ The gap anisotropy leads to a strong decreasing of the energy of the
flapping mode at lowering of the temperature, and $\omega _{s}$ tends to
zero when $T\rightarrow 0$, as shown in Fig. \ref{fig4}.

Near the pole $\Omega \simeq \Omega _{s}$ we can approximate%
\begin{align}
& \Pi _{\mathsf{weak}}^{\mu \nu }\left( k_{\mu }k_{\nu }-k^{2}g_{\mu \nu
}\right)   \notag \\
& \simeq \frac{-2\rho \left( \chi -\frac{3}{2}\Omega ^{2}\left\langle \bar{b}%
^{2}\mathcal{I}_{0}\right\rangle \right) \left\langle \left( 2\omega
^{2}-2q_{\parallel }^{2}-q_{\perp }^{2}\right) \bar{b}^{2}\mathcal{I}%
_{0}\right\rangle }{\Omega \left\vert \left( 1+\frac{4}{3}\mathfrak{g}%
_{0}\left\langle \bar{b}^{2}\mathcal{I}_{0}\right\rangle \right) \partial
\chi /\partial \Omega ^{2}-2\mathfrak{g}_{0}\left\langle \bar{b}^{2}\mathcal{%
I}_{0}\right\rangle ^{2}\right\vert _{\Omega =\Omega _{s}}\left( \Omega
-\Omega _{s}+i0\right) }.  \label{exp}
\end{align}%
The displacement of the pole in a complex $\omega $-plane is chosen so that
to obtain the retarded polarization. Such a displacement is equivalent to the
presence of a delta-function imaginary part%
\begin{align}
& \operatorname{Im}\Pi _{\mathsf{weak}}^{\mu \nu }\left( k_{\mu }k_{\nu
}-k^{2}g_{\mu \nu }\right)   \notag \\
& \simeq -\frac{2\pi \rho \left( \frac{3}{2}\Omega ^{2}\left\langle \bar{b}%
^{2}\mathcal{I}_{0}\right\rangle -\chi \right) \left\langle \left( 2\omega
^{2}-2q_{\parallel }^{2}-q_{\perp }^{2}\right) \bar{b}^{2}\mathcal{I}%
_{0}\right\rangle }{\Omega _{s}\left\vert \left( 1+\frac{4}{3}\mathfrak{g}%
_{0}\left\langle \bar{b}^{2}\mathcal{I}_{0}\right\rangle \right) \partial
\chi /\partial \Omega ^{2}-2\mathfrak{g}_{0}\left\langle \bar{b}^{2}\mathcal{%
I}_{0}\right\rangle ^{2}\right\vert _{\Omega =\Omega _{s}}}\delta \left(
\Omega -\Omega _{s}\right) .  \label{im}
\end{align}%
To evaluate this expression we expand the function $\mathcal{I}_{0}$ near
the pole to obtain%
\begin{equation}
\mathcal{I}_{0}\simeq \tilde{I}_{0}+\left( \Omega ^{2}-\Omega
_{s}^{2}\right) \tilde{I}_{1}  \label{I0}
\end{equation}%
with $\tilde{I}_{0}$, as given in Eq. (\ref{It}), and%
\begin{equation}
\tilde{I}_{1}\equiv \frac{\Delta ^{4}}{4}\int_{-\infty }^{\infty }\frac{%
d\varepsilon }{E^{5}}\tanh \frac{E}{2T}.  \label{I1t}
\end{equation}%
In this way we find%
\begin{equation}
\partial \chi /\partial \Omega ^{2}\simeq \frac{3}{4}\left( 2\left\langle
\allowbreak \left( 1+n_{3}^{2}\right) \tilde{I}_{0}\right\rangle
-\left\langle n_{3}^{2}\left( 1-n_{3}^{2}\right) \tilde{I}_{1}\right\rangle
\right)   \label{dhi}
\end{equation}%
and%
\begin{equation*}
\chi \left( \Omega _{s},y\right) \simeq \frac{1}{4}\left[ 6\Omega
_{s}^{2}\allowbreak \left\langle \left( 1+n_{3}^{2}\right) \tilde{I}%
_{0}\right\rangle -~3\left\langle n_{3}^{2}\left( 1-n_{3}^{2}\right) \tilde{I%
}_{0}\right\rangle +\left\langle \left( 1-3n_{3}^{2}\right) A\right\rangle %
\right] .
\end{equation*}%
The remaining calculations, similar to ones performed in the previous
section, result in the following neutrino energy losses caused by the decay
of spin waves:%
\begin{equation}
\epsilon _{\mathsf{SWD}}=\varepsilon _{0}\frac{64}{3\pi }\sqrt{6}\frac{\tau
^{7}y^{7}\Omega _{s}^{5}}{\exp \left( 2y\Omega _{s}\right) -1}\frac{\frac{3}{%
4}\left( \Omega _{s}^{2}\left\langle \bar{b}^{2}\tilde{I}_{0}\right\rangle -%
\frac{2}{3}\chi \right) \left\langle \bar{b}^{2}\tilde{I}_{0}\right\rangle }{%
\left( 1+\frac{4}{3}\mathfrak{g}_{0}\left\langle \bar{b}^{2}\tilde{I}%
_{0}\right\rangle \right) \partial \chi /\partial \Omega ^{2}-2\mathfrak{g}%
_{0}\left\langle \bar{b}^{2}\tilde{I}_{0}\right\rangle ^{2}}  \label{swd}
\end{equation}%
In this expression the spin-wave relative frequency $\Omega _{s}\left( \tau
\right) $ is defined by Eq. (\ref{omegas}). The SWD neutrino emissivity, as
obtained earlier in the average-angle approximation, can be obtained from
this expression if to replace the relative frequency with a constant $\Omega
_{s}\rightarrow 1/20$. 

It was already mentioned the gap anisotropy leads to a strong decreasing of
the wave energy along with lowering of the temperature. This is to suppress
substantially the neutrino energy losses caused by the spin wave decays because the rate of neutrino losses is strongly dependent on the wave energy.
\begin{figure}[h]
\includegraphics{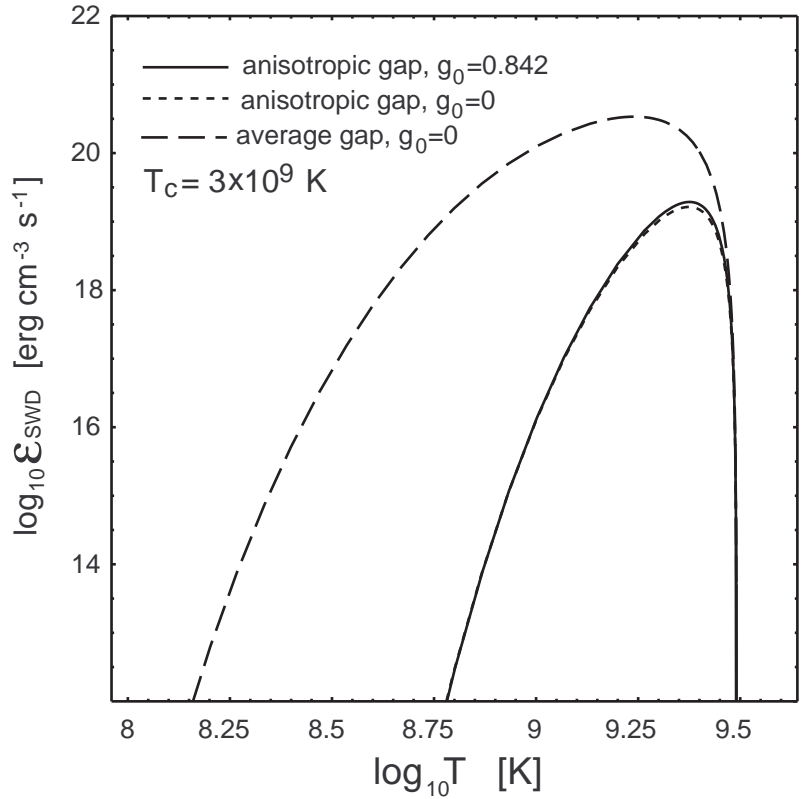}
\caption{SWD neutrino emissivity from neutron matter versus the temperature $%
T$ in logarithmic scale. The solid curve is calculated by Eq. (\protect\ref%
{eps2}) with taking into account the anisotropy of the energy gap and the
residual particle-hole interactions in the spin-spin channel with the Landau
parameter $\mathfrak{g}_{0}=0.842$. Short-dashed curve is same but with $%
\mathfrak{g}_{0}=0$. The long-dashed curve is calculated in the
average-angle approximation and with $\mathfrak{g}_{0}=0$, as given by Eq. (%
\protect\ref{epsav}).}
\label{fig5}
\end{figure}

 In Fig. \ref{fig5}, we show SWD neutrino emissivity from neutron
matter versus the temperature $T$ in logarithmic scale. We show also the
neutrino emissivity as calculated in the anisotropic BCS approximation and
with taking into account of both the anisotropy of the energy gap and
residual particle-hole interactions. The Fermi-liquid effects lead to a
minor (about 15\%) increase of the neutrino emissivity in the SWD channel,
however this emissivity is small in comparison with the result obtained in
the average-angle approximation

\subsection{Competitive neutrino processes}

\begin{figure}[h]
\includegraphics{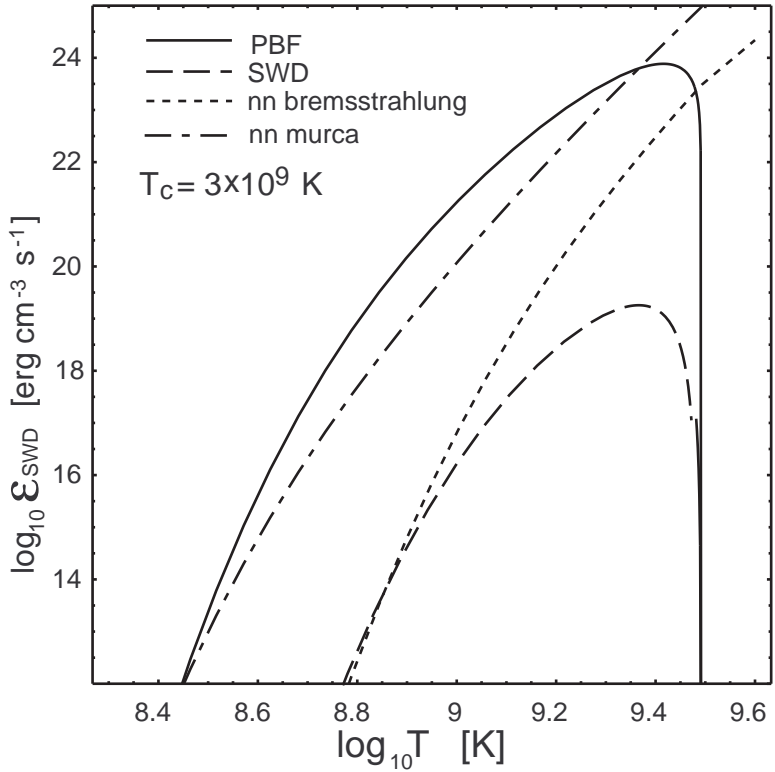}
\caption{PBF and SWD neutrino emissivities versus temperature $T$ in
comparison with the modified urca and bremsstrahlung emissivities at $k_{F}=1.7$.}
\label{fig6}
\end{figure}
The neutrino emissivity for the processes discussed above is plotted in Fig. %
\ref{fig6}, together with the modified urca and bremsstrahlung emissivities
with the suppression factors resulting from superfluidity as obtained in Ref. \cite{YKGH}.

The emissivity from the PBF dominates everywhere below the critical
temperature for the $^{3}$P$_{2}$ superfluidity except the narrow
temperature domain near the critical point, where the modified urca
processes are more operative. Neutrino losses caused by the bremsstrahlung
and SWD are less effective than in the PBF processes.

\section{Summary and conclusion}

We have calculated the neutrino energy losses from the $^{3}P_{2}(m_{j}=0)$
superfluid neutron liquid with accurate taking into account of the
anisotropy of the superfluid energy gap and minimal account of residual
Fermi-liquid interactions. In our analysis we have used the integral
expressions for anomalous three-point effective vertices derived in Refs. 
\cite{L10a,L12}. We have examined the neutrino energy losses through neutral
weak currents caused by the pair breaking and formation processes and by the
spin-wave decays. The corresponding neutrino emissivity in is given in Eqs. (%
\ref{eps2}) and (\ref{swd}).

Earlier the neutrino losses have been calculated discarding the Fermi-liquid
effects with using the average-angle approximation, where the anisotropic
energy gap in equations for the anomalous vertices was replaced by its
average-angle magnitude, $\bar{b}^{2}\Delta ^{2}\left( T\right) \rightarrow
\Delta ^{2}\left( T\right) $. A comparison of the results allows to make
some inferences concerning validity of the average-angle approach.

As shown above, the gap anisotropy leads to quenching of the PBF neutrino
emissivity on about 15\%, while the residual particle-hole interactions
increase the PBF emissivity on about 5\%. This difference is practically
indistinguishable in the logarithmic scale in Fig. \ref{fig6}. Up to
accuracy about ten percents, one can neglect both the anisotropy of the
effective weak vertices and the Fermi-liquid effects in the PBF processes
and use the simple expression in the form of one-fold integral (\ref{epsav})
for practical estimates.

However, the exact account of the anisotropy dramatically modifies the SWD
neutrino losses. This fact has a simple explanation. As found in Ref. \cite%
{L10a} , the relative spin-wave energy is constant in the average-angle
approximation, $\Omega _{s}^{\mathsf{av}}=1/\sqrt{5}$. Taking into account
of the gap anisotropy leads to a strong temperature dependence of the energy
of this collective mode \cite{L12}. The wave frequency diminishes and tends
to zero at the lowering of the temperature, as shown in Fig. \ref{fig4}.
Since the rate of the neutrino losses is strongly dependent on the wave
energy the anisotropy leads to a rapid decrease of the SWD energy losses
along with lowering of the temperature. A comparison of different
competitive processes presented in Fig. \ref{fig6} allows to conclude that
the dominant energy losses from the $^{3}P_{2}$ superfluid neutron liquid
are generated by the PBF processes except the narrow temperature domain near
the critical point, where the modified urca processes are more operative.

\begin{appendix}

\section{Neutrino emissivity in the averege-angle approximation}

\label{sec:A}The neutrino emissivity, as obtained in Ref. \cite{L10a} in the
averege-angle approximation, is written in the form of the two-fold integral%
\begin{equation}
\epsilon _{\mathsf{av}}\simeq \frac{2}{15\pi ^{5}}G_{F}^{2}C_{\mathsf{A}}^{2}%
\mathcal{N}_{\nu }p_{F}M^{\ast }T_{c}^{7}\tau ^{7}y^{2}\int \frac{d\mathbf{n}%
}{4\pi }\bar{b}^{2}\int_{0}^{\infty }dx\frac{z^{4}}{\left( 1+e^{z}\right)
^{2}}~,  \label{eps}
\end{equation}%
where $z=\sqrt{x^{2}+\bar{b}^{2}y^{2}}$. With the aid of the change $x=y%
\sqrt{\Omega ^{2}-\bar{b}^{2}}$ $\allowbreak $ one can recast this
expression to the form%
\begin{equation}
\epsilon _{\mathsf{av}}\simeq \frac{2}{15\pi ^{5}}G_{F}^{2}C_{\mathsf{A}}^{2}%
\mathcal{N}_{\nu }p_{F}M^{\ast }T_{c}^{7}\tau ^{7}y^{2}\int \frac{d\mathbf{n}%
}{4\pi }\bar{b}^{2}\int_{0}^{\infty }dx\frac{\left( x^{2}+\bar{b}%
^{2}y^{2}\right) ^{2}}{\left( 1+e^{y\Omega }\right) ^{2}}  \label{eps4}
\end{equation}
Further simplification is possible if to change the order of
integration and
write the emissivity in the form%
\begin{equation}%
\epsilon _{\mathsf{av}}\simeq \frac{2}{15\pi ^{5}}G_{F}^{2}C_{\mathsf{A}}^{2}\mathcal{N}_{\nu }p_{F}M^{\ast }T_{c}^{7}\tau
^{7}y^{7}\int_{0}^{\infty
}d\Omega \frac{\Omega ^{5}}{\left( 1+e^{y\Omega
}\right) ^{2}}\int \frac{d%
\mathbf{n}}{4\pi }\bar{b}^{2}\frac{\Theta
\left( \Omega -\bar{b}\right) }{%
\sqrt{\Omega ^{2}-\bar{b}^{2}}},  \label{%
eps5}
\end{equation}%
where $\bar{b}^{2}=\left( 1+3n_{3}^{2}\right) /2$
and $\Theta \left(
x\right) $ is the Heavyside's step-function. The
integation over the Fermi
surface gives%
\begin{align}
\int \frac{d%
\mathbf{n}}{4\pi }\bar{b}^{2}\frac{\Theta \left( \Omega -\bar{b}%
\right) }{%
\sqrt{\Omega ^{2}-\bar{b}^{2}}}& =\frac{\pi }{4\sqrt{6}}\left(
1+2\Omega
^{2}\right) \Theta \left( \Omega -1/\sqrt{2}\right) \Theta \left( 
\sqrt{2}%
-\Omega \right)  \notag \\
& +\frac{\sqrt{6}}{12}\left( \left( 2\Omega
^{2}+1\right) \arcsin \frac{%
\sqrt{3}}{\sqrt{2\Omega ^{2}-1}}-\sqrt{%
6\Omega ^{2}-12}\right) \Theta \left(
\Omega -\sqrt{2}\right)  \label{dn}
\end{align}%
Inserting this expression into Eq. (\ref{eps5}) we arrive to
Eq. (\ref{epsav}%
).
\end{appendix}

\end{document}